\documentclass[twocolumn,showpacs,preprintnumbers,prl]{revtex4}

\usepackage{graphicx}

\usepackage{dcolumn}

\usepackage{bm}

\begin{document}

\title{Spin-transfer torques in anti-ferromagnetic metals from first principles}

\author{Yuan Xu}

\author{Shuai Wang}

\author{Ke Xia}

\affiliation{State Key Laboratory for Surface Physics, Institute
of Physics, Chinese Academy of Sciences, P.O. Box 603, Beijing
100080, China }

\date{\today}

\begin{abstract}

In spite of the absence of a macroscopic magnetic moment, an
anti-ferromagnet is spin-polarized on an atomic scale. The electric
current passing through a conducting anti-ferromagnet is polarized
as well, leading to spin-transfer torques when the order parameter
is textured, such as in anti-ferromagnetic non-collinear spin valves
and domain walls. We report a first principles study on the
electronic transport properties of anti-ferromagnetic systems. The
current-induced spin torques acting on the magnetic moments are
comparable with those in conventional ferromagnetic materials,
leading to measurable angular resistances and  current-induced
magnetization dynamics. In contrast to ferromagnets, spin torques in
anti-ferromagnets are very nonlocal. The torques acting far away
from the center of an anti-ferromagnetic domain wall should
facilitate current-induced domain wall motion.

\end{abstract}

 \maketitle










Anti-ferromagnet metals (AFMs) are widely used to pin and
exchange-bias ferromagnets in devices such as magnetic spin valve
read heads \cite{exchangebias,EBSV_Bass}. The absence of magnetic
stray fields makes them very useful as probe tips in spin-polarized
scanning tunnelling microscopes \cite{Cr_STM_tip}. AFMs are
materials with spontaneous magnetization below a critical N\'{e}el
temperature at which the magnetic moments of two (or more)
sub-lattices point into opposite directions, such that the net
magnetic moment of AFM vanishes. External magnetic fields cannot be
used to manipulate the strongly coupled magnetic moments of AFMs. An
alternative for magnetic fields to excite ferromagnets (FMs) is the
current-induced spin-transfer torque (STT) \cite{torques}. Since the
electric current is spin-polarized on an atomic scale, the STT
phenomenon may excite the anti-ferromagnetic order parameter. It is
not obvious whether the STT is strong enough and how different
sub-lattices are affected.

Recent theoretical studies
\cite{AFSV_MacDonald,FM_AFSV_MacDonald,AuCr_MarDonald} predicted
that in spin valves with uncompensated anti-ferromagnets angular
resistance (AR) effect can be expected. Reversal of the
anti-ferromagnetic order parameter is equivalent with a phase shift
by a half period. Furthermore, STT is not an interface effect as in
ferrromagnets, but effective over the whole AFM. Very recently,
strong experimental evidences of STT between AFM and FM have been
reported \cite{EBSV_Tsoi}. In addition, the resistance of
anti-ferromagnetic domain walls (AFDWs) in chromium has been
measured\cite{transport_AF_Cr_domain_wall}.

In contrast to FMs, AFMs have much smaller shape anisotropies which
makes manipulation of its magnetic moments easier --- provided one
can apply a \emph{sizable} spin torque on the moments. What is the
magnitude of the spin torque for an atomic scale spin polarization?
Can an atomic scale spin polarization gives rise to observable AR
effect? These are the questions we attempt to answer in this work.

$\gamma$-FeMn is an AFM used as pinning layer in spin valves, and we
demonstrate that an electric current can induce angular momentum
transfer between magnetic moments in AFMs. By a detailed analysis of
the current and non-equilibrium charge densities, we show that an
atomic scale spin polarization exists even for compensated AFMs. A
sizable AR is predicted in anti-ferromagnetic spin valves (AFSVs).
In an AFDW the spin torque is effective not only around the domain
wall center but throughout the entire AFM, which can be understood
by the suppressed spin precession.


In our calculations, the atomic potential was determined in the
framework of the tight-binding (TB) linear muffin-tin-orbital (MTO)
method based on density functional theory in the local density
approximation \cite{Disorder_book} and an exchange-correlation
potential parameterized by von Barth and Hedin
\cite{EX_Barth_Hedin}. The self-consistent crystal potentials were
used as input to a TB-MTO wave-function-matching calculation and the
scattering wave functions of the whole system were obtained
explicitly \cite{master}. To model non-collinear magnetic
configurations in AFDWs, a rigid potential approximation is applied,
which is a good approximation for sufficiently wide domain walls
\cite{justification_rotate_H}.

\begin{figure}[tbh]
\includegraphics[width=6.5cm, bb=10 9 584 209]{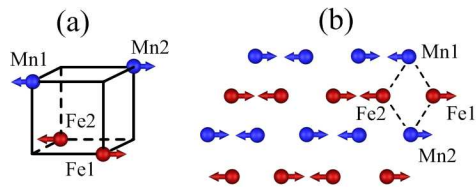}
\caption{(color online) (a) Magnetic structure of
$\protect\gamma$-FeMn. (b) $fcc$(111) plane of
$\protect\gamma$-FeMn, where the dashed lines connect the atoms in a
unit cell.} \label{fig:3}
\end{figure}

The spin current and charge density can be calculated from the
scattering wave functions. The torques acting on an atomic plane
are defined as the difference between incoming and outgoing spin
current \cite{master_Wang}. Finally, we focus on transport along
the direction perpendicular to the plane $fcc(111)$ as illustrated
in Fig.\ref{fig:3}(b), where four atoms connected by the dashed
lines form a unit cell. $fcc(111)$ is a compensated plane with
zero net magnetic moment in each layer.

The magnetic structure of $\gamma$-FeMn is shown in
Fig.\ref{fig:3}(a). This configuration has been observed
experimentally \cite{Kouvel_collinear_FeMn} and confirmed to be
energetically favorable by \emph{ab.initio} calculation
\cite{Hafner_FeMn}. There are four inequivalent atoms in a unit
cell. The magnetic moments of Fe2 and Mn1 and those of Fe1 and Mn2
point into opposite directions. Our calculation
\cite{AS_sphere_FeMn} yields the magnetic moments as
$m_{Fe}=1.4\mu_{B}$ and $m_{Mn}=1.9\mu_{B}$.

\begin{figure}[tbp]
\includegraphics[width=8cm, bb=21 17 325 227]{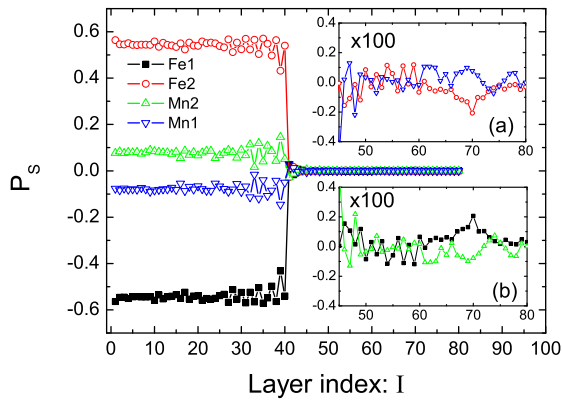}
\caption{(color online) Layer resolved spin-polarization of
non-equilibrium charge density, $P_{S}^{\alpha}$, on the four
inequivalent sub-lattices of the FeMn$|$Cu system, where FeMn
occupies 1st-40th layers and Cu 41th-80th layers. The electron
current flows from FeMn to Cu. Inset (a) gives the $P_{S}$ on the Cu
side for the inequivalent sub-lattices corresponding to Fe2 and Mn1,
Inset (b) gives the $P_{S}$ on the Cu side for the inequivalent
sub-lattices corresponding to Fe1 and Mn2.} \label{fig:13}

\end{figure}

When a current passes through an anti-ferromagnetic material, the
current-induced non-equilibrium charge density integrated over the
atomic sphere of an atom is spin-polarized. We define the
polarization of atom $\alpha$ as $P_{S}^{\alpha}=
(\rho_{\uparrow}^{\alpha}-\rho_{\downarrow}^{\alpha})/(\rho_{\uparrow}^{\alpha}+
\rho_{\downarrow}^{\alpha})$, where
$\rho^{\alpha}_{\uparrow(\downarrow)}$ is the spin resolved
non-equilibrium charge density. $\uparrow(\downarrow)$ indicates the
electron spin parallel or antiparallel to the spin quantization
axis.

In Fig.\ref{fig:13} we plot $P_{S}^{\alpha}$ for a single FeMn$|$Cu
interface. We group the system into four inequivalent sub-lattices
according to the inequivalent atoms in the unit cell of
$\gamma$-FeMn. A nonzero polarization is found even at the Cu side.
The sum of the spin polarizations of the four sub-lattices vanishes
in each layer. These results show that a compensated
anti-ferromagnet injects a spin polarization into a normal metal
that oscillates on an atomic scale.

Let us consider now an AFSV denoted as
FeMn1($\theta$)$|$Cu$|$FeMn2$|$Cu, where an infinitely thick FeMn1
serves as the fixed layer and FeMn2 as the free layer. The $z$ axis
of spin space is set along the magnetization of the free layer and
$\theta$ gives the relative angle between both order parameters. In
our calculations, the thicknesses of the Cu spacer and the FeMn2
free layer are chosen to be both 10 monolayers (ML)
 \cite{FeMn_111_plane_shift}.

\begin{figure}[tbp]
\includegraphics[width=7.5cm]{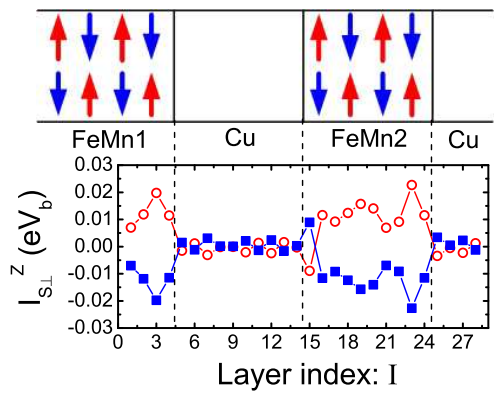}
\caption{(color online) Top panel: cartoon of the parallel
configuration, $\theta=0^{o}$, of an AFSV. The red and blue arrows
denote the two sub-lattices A and B, respectively. Bottom panel: the
layer resolved $z$ component of spin-polarized current,
I$_{s\bot}^{z}$, where the subscript $\bot$ denotes the current
perpendicular to the fcc(111) plane and the red circles (blue
squares) represent the spin-polarized current on the sub-lattice A
(B).} \label{fig:17}
\end{figure}

The top panel of Fig.\ref{fig:17} illustrates the magnetic
structure of the so-called parallel configuration
($\theta=0^{o}$). Here and in the following we group the four
sub-lattices into two according to the direction of their magnetic
moments. When the magnetization of FeMn1 is rotated from
$\theta=0^{o}$ to $\theta=180^{o}$ (the antiparallel
configuration), the electric resistance changes.

The bottom panel of Fig.\ref{fig:17} gives the spin-polarized
current at the two sub-lattices. When the magnetizations of the two
AFM layers are collinear, there is no net current spin polarization.
However, the spatial distribution of spin-polarized current is not
trivial. In fact, the spin-polarized currents on the two
sub-lattices do not vanish even in Cu. They are identical in
magnitude but oppositely spin-polarized. Another finding is that the
spin-polarized current oscillates between the two sub-lattices and
is not conserved within one sub-lattice.

\begin{figure}[tbp]
\includegraphics[width=8cm, bb=21 17 325 227]{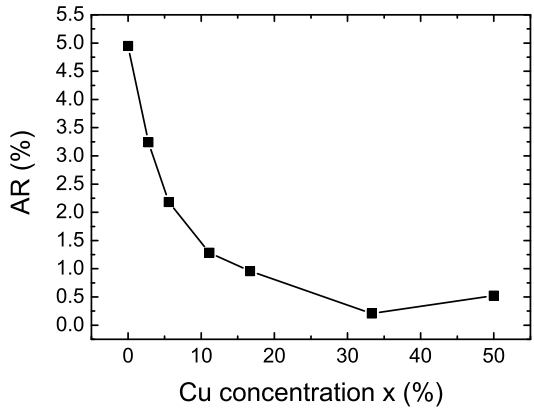}
\caption{Cu concentration $x$ dependence of AR in the system
FeMn1$|$Cu(10ML)$| $FeMn2(10ML)$|$Cu.} \label{fig:4}
\end{figure}

The AR ratio in terms of conductance as
$[G(0^{o})-G(180^{o})]/G(180^{o})$ amounts to a measurable $5\%$.
The remaining question is the robustness of this AR effect in the
presence of interfacial disorder. We modelled disorder by an
interfacial alloy with a lateral super-cell method \cite{master}.
For a 1ML (FeMn)$_{1-x}$Cu$_{x}$ interfacial alloy, the AR
dependence on interfacial alloy concentration $x$ is shown in
Fig.\ref{fig:4}, which indicates that the AR is somewhat suppressed
in the presence of disorder, but should remain to be observable.

It is interesting to investigate the STT due to the atomic scale
spin dependent scattering. Before going into the numerical results,
we analyze the symmetry of the STT. For one sub-lattice, similar to
that of FM, the torque on a local magnetic moment $\mathbf{m}$ is
proportional to $\mathbf{m}\times\mathbf{M}\times\mathbf{m}$
(in-plane) and $\mathbf{m}\times\mathbf{M}$ (out-of-plane), where
$\mathbf{M}$ denotes the source of spin current. For another
sub-lattice, both $\mathbf{m}$ and $\mathbf{M}$ are reversed. The
in-plane torque is also reversed whereas the out-of-plane torque
remains unchanged. Because the in-plane torque
$\mathbf{T}_{\parallel}$ on two sub-lattices are identical in
magnitude and opposite in direction and the magnetization of two
sub-lattices points into opposite directions, the net effect of STT
tends to rotate those moments together. In contrast to
$\mathbf{T}_{\parallel}$, the out-of-plane torque
$\mathbf{T}_{\perp}$ acts identically on the two magnetic
sub-lattices. As the exchange coupling in the AFM is strong,
out-of-plane torques will not result in any significant effect on
the dynamics of the magnetic moments.

\begin{figure}[tbp]
\includegraphics[width=8cm, bb=12 10 224 247]{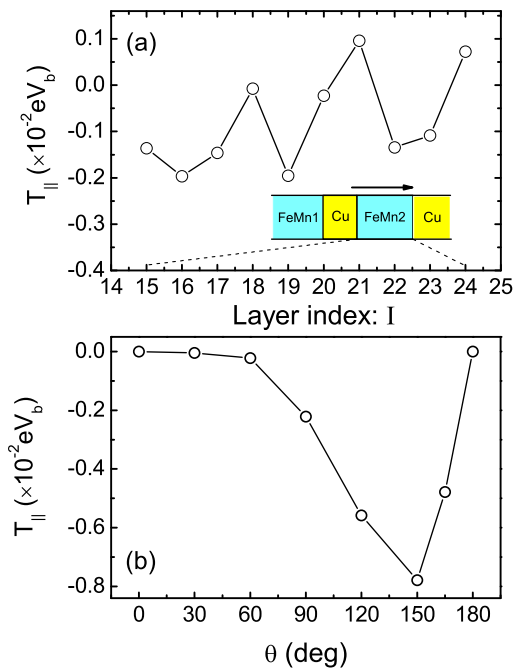}
\caption{(color online) (a) The layer resolved in-plane torque
$T_{\parallel}$ on one sub-lattice (Fe1$\&$Mn2) of FeMn2 in
FeMn1($\protect\theta$)$|$Cu(10ML)$| $FeMn2(10ML)$|$Cu, as shown by
the cartoon. Here $\protect\theta=150^{o}$, electrons flow from
FeMn1 to FeMn2 and FeMn2 is located at $15\leq I\leq 24$. (b)
Angular dependence of the in-plane torque $T_{\parallel}$ on the
sub-lattice (Fe1$\&$Mn2) of FeMn2 in the same system.} \label{fig:7}
\end{figure}

As shown in Fig.\ref{fig:7}(a), the spin torque in the AFM layer
oscillates deep into material which differs from conventional FM
spin valve \cite{Spin_torque_Stile,Spin_torque_Zwierzycki}. The
angular dependence of the total in-plane torque acting on
sub-lattice (Fe1$\&$Mn2) in FeMn2 is shown in Fig.\ref{fig:7}(b).
For $\theta=150^{o}$, the torque ($4.1\times10^{-4}eV_{b}/\mu_{B}$)
acting on the surface atoms is much smaller than that in FM spin
valve ($\sim 10^{-3}eV_{b}/\mu_{B}$) \cite{Edwards05}. However, due
to slow decay of the spin torque in AFMs, the integrated torque on
per atom of AFMs ($2.4\times10^{-4}eV_{b}/\mu_{B}$) is comparable to
that in FMs.

\begin{figure}[tbp]
\includegraphics[width=8cm, bb=11 11 213 217]{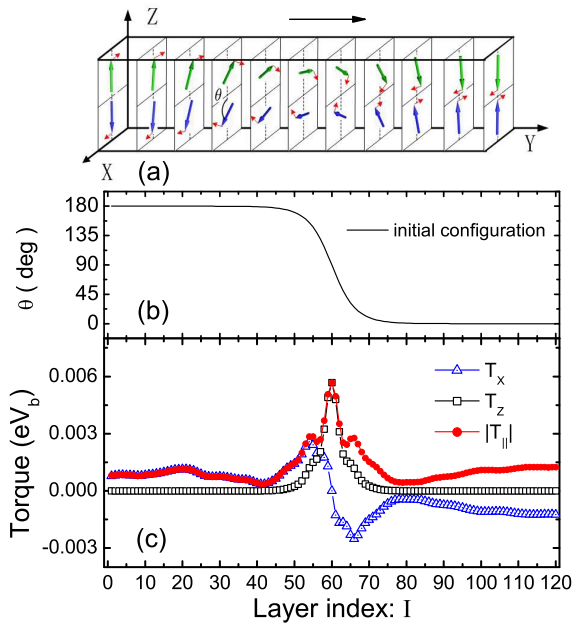}
\caption{(color online) (a) Schematic view of the AFDW, in which two
sub-lattices are indicated by the green and blue arrows, the
magnetization is rotated in the plane by an angle $\protect\theta$
and the electrons are injected from the left side. The in-plane
torque directions are indicated by the red arrows. (b)
$\protect\theta$ as the function of position. (c) Spin torques
acting on the sub-lattice (Fe1$\&$Mn2) in each monolayer with
initial configuration, where $T_{X}$, $T_{Z}$ denote the in-plane
components and $|T_{||}|$ is the absolute value of in-plane
torques.} \label{fig:11}
\end{figure}

Similar spin transfer effect should occur in AFDWs. As domain wall
physics in AFMs has not been fully understood
\cite{DWconfigurations}, we simply consider a Bloch-like domain wall
which could exist in the experiment \cite{Zabel_exchange_bias}, as
shown in Fig.\ref{fig:11}(a). The configuration is described by
$\theta(y)=\frac{\pi}{2}+\arcsin[\tanh(\frac{y-y_{o}}{\lambda_{DW}})]$,
where $\lambda_{DW}$ is the characteristic length which is selected
to be 4ML and $y_{o}$ the center of the wall. The spin direction in
each monolayer along the wall is shown in Fig.\ref{fig:11}(b). The
change of the total conductance due to the formation of a wall is
2.8$\%$ for a 4ML thick domain wall, which is comparable to that in
a FM with the same width.

Fig.\ref{fig:11}(c) shows the in-plane torque exerted on one
sub-lattice (Fe1$\&$Mn2) as a function of position in the domain
wall. There is a substantial STT in the region far away from the
domain wall center. This is attributed to the spin degeneracy of the
electronic band structure in the AFM. The injected spin can
penetrate and precess coherently until deep into the AFM. As a
result, STT in AFM is very non-local, at least in the absence of
disorder scattering.

We investigate the dynamics of the AFDW with the
Landau-Lifshitz-Gilbert (LLG) equation \cite{Zhang_domainwall} using
our calculated STT. An AFM with two sub-lattices is described by two
sets of coupled LLG equations. Here we decouple the equations by
assuming that the magnetizations of two sub-lattices in the same
layer are antiparallel to each other at any time. This should be
reasonable due to a strong anti-ferromagnetic coupling between the
two sub-lattices.

To investigate dynamics of the domain wall in the presence of a
current, we follow the method of Ohe and Kramer \cite{ohe} who
studied FM domain walls. This method consists of several steps: (i)
Given a domain wall configuration, we calculate the STT acting on
each site under a certain current density; (ii) the obtained spin
torque at each site is substituted into the LLG equation to obtain a
set of equations for each site in the domain wall at a given time
$t$; (iii) by integrating the LLG equation, we obtain the domain
wall configuration after an interval at $t'=t+\delta t$; (iv) once a
new domain wall configuration is obtained, we can calculate the STT
again and the process is repeated. The interval $\delta t$ between
two consecutive steps is chosen small enough to guarantee numerical
convergence of the solution.

In particular, a magnetic anisotropy constant of
$1.35\times10^{5}$erg/cm$^{3}$, exchange stiffness constant of
$0.94\times10^{-9}$ erg/cm, magnetization of sub-lattice of $650$ Gs
and Gilbert damping constant of $0.1$ are used in our calculation.

Under a current density of $5\times10^{7}$ A/cm$^{2}$ and starting
from the initial domain wall configuration mentioned above, we
obtained results for 10 time steps of $\delta t= 2$ ps. The domain
wall moves into the $-Y$ direction and the estimated velocity is 4
m/s. In contrast to the FM domain wall, the velocity is maintained
due to the vanishing de-magnetization field. For the same reason,
when no pinning field is taken into account, the critical current
that sets the AFDW into motion is also expected to be very small as
compared to a FM domain wall.


In summary, we find an atomic scale spin polarization in completely
compensated AFMs. An STT can be induced by an electric bias. In
AFSVs, a sizable AR is predicted and the STT is found to be
comparable to that in conventional FM spin valves. In AFDWs, the
torques turn out to be nonlocal. The spin dynamics of AFDWs is
studied by including the STT into the LLG equation for AFMs.

We acknowledge G. E. W. Bauer and H. Guo for their critical
reading of the manuscript. We also acknowledge M. D. Stiles for
his suggestion to define an atomic scale spin polarization. This
work is supported by the NSF(10634070) and MOST(2006CB933000,
2006AA03Z402) of China.


\begin{references}



\bibitem{exchangebias} W. H. Meiklejohn and C. P. Bean, Phys. Rev. {\bf 102}, 1413
(1956).


\bibitem{EBSV_Bass}J. Bass, A. Sharma, Z. Wei, M. Tsoi, arXiv:0804.3358


\bibitem{Cr_STM_tip}A. Kubetzka, {\it et al.},
Phys. Rev. Lett. {\bf 88}, 057201 (2002).

\bibitem{torques}J.C. Slonczewski, J. Magn. Magn. Mater. {\bf 159},
L1 (1996). L. Berger, Phys. Rev. B {\bf 54}, 9353 (1996). J.A.
Katine, {\it et al.}, Phys. Rev. Lett. {\bf 84}, 3149 (2000);
X. Waintal, {\it et al.},
Phys. Rev. B {\bf 62}, 12317 (2000). J. Z. Sun, J. Magn. Magn.
Mater. {\bf 202}, 157 (1999). M. Tsoi {\it et al.}, Phys. Rev. Lett.
{\bf 80}, 4281 (1998).



\bibitem{AFSV_MacDonald}A. S. N\'{u}\~{n}ez, {\it et al.},
Phys. Rev. B {\bf 73}, 214426 (2006).

\bibitem{FM_AFSV_MacDonald}R. A. Duine, {\it et al.},
Phys. Rev. B {\bf 75}, 014433 (2007).

\bibitem{AuCr_MarDonald}P. Haney, {\it et al.},
Phys. Rev. B {\bf 75}, 174428 (2007).

\bibitem{EBSV_Tsoi}Z. Wei, {\it et al.},
Phys. Rev. Lett. {\bf 98}, 116603 (2007).


\bibitem{transport_AF_Cr_domain_wall}R. Jaramillo, {\it et al.},
Phys. Rev. Lett. {\bf 98}, 117206 (2007).

\bibitem{Disorder_book}I. Turek, {\it et al.}
{\it Electronic Structure of Disordered Alloys, Surfaces and
Interfaces} (Kluwer, Boston-London-Dordrecht, 1997).

\bibitem{EX_Barth_Hedin}U. von Barth and L. Hedin, J. Phys. C {\bf 5}, 1629 (1972).


\bibitem{master}K. Xia {\it et al.}, Phys. Rev. B {\bf 73}, 064420
(2006).

\bibitem{justification_rotate_H}Selfconsistent calculations show that when the relative angle
between neighboring magnetic atoms is less than $20^{o}$, magnetic
moments change less than $2\%$.

\bibitem{master_Wang}S. Wang, Y. Xu, and K. Xia, arXiv:0801.3135.


\bibitem{Kouvel_collinear_FeMn}J. S. Kouvel and J. S. Kasper, J. Phys. Chem. Solids {\bf 24}, 529
(1963), P. Bisanti, G. Mazzone and F. Sacchetti, J. Phys. F: Met.
Phys {\bf 17}, 1425 (1987).






\bibitem{Hafner_FeMn}D. Spi\v{s}\'{a}k and J. Hafner, Phys. Rev. B {\bf 61}, 11569 (2000).

\bibitem{AS_sphere_FeMn}The Wigner-Seitz radii of the two atoms satisfy $r_{Fe}/r_{Mn}=0.98$. We use a grid of 57600 $k_{\|}$ points in the full 2DBZ for the transport calculations.

\bibitem{FeMn_111_plane_shift}Several relative configurations between the two FeMn (111) planes facing each other across the
Cu spacer can be considered. In this work, we take the most
convenient one by replacing 10 ML of the bulk FeMn with Cu atoms.

\bibitem{Spin_torque_Stile}M. D. Stiles, {\it et al.}, Phys. Rev. B {\bf 66}, 014407 (2002).




\bibitem{Spin_torque_Zwierzycki}M. Zwierzycki, {\it et.al.}, Phys. Rev. B {\bf 71}, 064420 (2005).



\bibitem{Edwards05}D. M. Edwards, {\it et.al.},
Phys. Rev. B {\bf 71}, 054407 (2005).

\bibitem{DWconfigurations} M. Ando and S. Hosoya, Phys. Rev. Lett. {\bf 29}, 281 (1972). F. Nolting {\it et.al.}, Nature (London) {\bf 405}, 767 (2000).
C. L. Chien, {\it et.al.},
Phys. Rev. B {\bf 68}, 014418 (2003).



\bibitem{Zabel_exchange_bias}F. Radu and H. Zabel, arXiv:0705.2055.



\bibitem{Zhang_domainwall}Z. Li and S. Zhang, Phys. Rev. B {\bf 70}, 024417 (2004).



\bibitem{ohe}J. Ohe and B. Kramer, Phys. Rev. Lett. 96, 027204(2006)















\end{references}
\end{document}